\documentclass[12pt,amsmath,amsfonts,amssymb,graphics,setspace]{article}

\def\half{\mbox{$\frac{1}{2}$}}

\usepackage{latexsym}

\newtheorem{prop}{Proposition}
\newtheorem{dfn}{Definition}
\newtheorem{dfnprop}{Definition-Proposition}

\newtheorem{lem}{Lemma}
\newtheorem{lem-theo}{Lemma-Theorem}
\newtheorem{cor}{Corollary}
  {\bigskip\noindent{\bf Acknowledgements}\bigskip}%
  {\bigskip}
\newcommand{\equal}{\!\!\!&=&\!\!\!}
\newcommand{\beqn}{\begin{equation}}
\newcommand{\eeqn}{\end{equation}}
\newcommand{\beqnarray}{\begin{eqnarray}}
\newcommand{\eeqnarray}{\end{eqnarray}}

\def\tanh{\mbox{\rm tanh}}
\def\sech{\mbox{\rm sech}}
\begin{document}
\date{ }

\title{A study of two new generalized negative KdV type equations} 

\author{Partha Guha\footnote{Email: $partha@bose.res.in$} \\ 
S.N. Bose National Centre for Basic Sciences\\
JD Block, Sector-3, Salt Lake \\ 
Kolkata-700098, India. \\
\and 
P G L Leach\footnote{E-mail leach@ucy.ac.cy}\,\,\footnote{Permanent address: Department of Mathematics and Institute of Sysyems Science, 
Durban University of Technology, PO Box 1334, Durban 4000, Republic of South Africa}\\
Department of Mathematics and Statistics,University of
Cyprus\\ Lefkosia 1678, Cyprus.\\
}

\maketitle

\vspace{.1in}
\hspace{1.10in}

\abstract{We give a simple geometric interpretation of the mapping of the negative KdV equation  
 $(- \frac{\psi_{xx}}{\psi})_t = (\psi^2)_x$ as proposed by Qiao and Li \{ arXiv:1101.1605 [math-ph], Europhys. Lett., 94 (2011) 50003\} and the Fuchssteiner equation $(\frac{u_{txx}}{u_x})_x + 4(\frac{uu_{t}}{u_x})_x + 2u_t = 0$ 
using geometry of projective connection on $S^1$ or stabilizer set of the Virasoro orbit.
We propose a similar connection between $(- \frac{\psi_{xx}}{\psi})_t = (\psi^3)_x$ and $(- \frac{\psi_{xx}}{\psi})_t = (\psi^4)_x$ with the higher-order negative KdV equations of Fuchssteiner type described as 
$(\frac{u_{txxx}}{3u^2 + u_{xx}})_x + 10(\frac{(uu_t)_x}{3u^2 + u_{xx}})_x + 3u_t = 0$ and
$\frac{u_{txxxx}}{(u^{\prime\prime\prime} + 16uu^{\prime}}\Big)_x + 20(\frac{uu_{txx}}
{(u^{\prime\prime\prime} + 16uu^{\prime}})_x + 30(\frac{u_xu_{tx}}{(u^{\prime\prime\prime} + 16uu^{\prime}})_x
 + 18(\frac{u_{xx}u_{t} + 64u^2u_t}{(u^{\prime\prime\prime} + 16uu^{\prime}})_x + 4u_t = 0$ respectively.
We study the Painlev\'e and symmetry analyses 
of these newly found equations and show that they yield soliton solutions. }

\bigskip


\bigskip


\section{Introduction}

It has been shown by Qiao and his collaborators (cf. \cite{Qiao,Qiao1}) that the following integrable system 
\beqn\label{1a}
\left(- \frac{\psi_{xx}}{\psi}\right)_t = 2\psi \psi_x,
\eeqn
is actually related to the first member of the  
negative KdV equation of Fuchssteiner type 
\beqn\label{1b}
\left(\frac{u_{txx}}{u_x}\right)_x + 4\left(\frac{uu_{t}}{u_x}\right)_x + 2u_t = 0.
\eeqn
Fuchssteiner \cite{Fuch} gave
a hodograph link from a B\"acklund transformation of the Camassa-Holm equation (CH)
\beqn\label{1c}
m_t+m_xu+2mu_x = 0,\qquad  \hbox{ where } m = u-u_{xx},
\eeqn
to this particular member of the (negative) KdV-hierarchy. Therefore the equation (\ref{1a}) yields a simpler reduced form of
the CH equation. The Lax pair of (\ref{1a}) is derived to guarantee its integrability.  
Furthermore the equation is shown to have classical solitons, periodic solitons and kink solutions.

\smallskip

Inspired by Fuchssteiner's work, Schiff \cite{Sch} introduced the associated Camassa-Holm (ACH) equation
and showed that it is related the Camassa-Holm equation.  He derived B\"acklund transformations by a 
loop group technique and used these to obtain some simple soliton and rational solutions. 
Hone \cite{Hone} showed that, because the hodograph transformation is essentially the same as 
in \cite{Fuch}, 
the ACH equation is naturally related to the inverse or negative KdV equation and has a Lax pair of which one part is 
just a (time-independent) Schr\"odinger equation. 

\smallskip

Qiao and Li \cite{Qiao} unified the positive- and negative-order KdV hierarchtes as
\beqn\label{1d}
v_{t_k} = JH_k = KH_{k-1}, \qquad \forall \, \, k \in {\bf Z}, 
\eeqn
where $K = \frac{1}{4}u^{-2}\partial u^2 \partial u^2 \partial u^{-2}$ and $J = \partial$.
The Hamiltonians $H_k$ are defined via recursion operators ${\cal R} = J^{-1}K$
and ${\cal R}^{-} = K^{-1}J$, which is given by $H_k ={\cal R}^kH_0$ and $H_k ={\cal R}^{k+1}H_{-1}$.  
This yields the entire KdV hierarchy, the positive order ($k \geq 0$)
gives the regular KdV hierarchy, while the negative
order ($k < 0$) produces some interesting equations gauge-equivalent to the Camassa-Holm
equation. This hierarchy possesses the bi-Hamiltonian structure because of
the Hamiltonian properties of $ J$. The second positive member of the hierarchy is the well-known KdV equation
$$
v_{t_2} = \frac{1}{2}v_{xxx} + \frac{3}{2}vv_{x},
$$
where $v = -u_{xx} / u$. When $k = -1$, this coincides with equation (\ref{1a}).

\smallskip

The stabilizer orbit of the coadjoint action
of the Virasoro algebra on its dual has been studied and it is known 
that many integrable ODEs are connected to this set \cite{Gu,Gu1}. 
It has been shown that by using Kirillov's superalgebra \cite{kiri,kiri1} it is possible to describe
the solution of the integrable systems associated to
the stabilizer orbit. We formulate the solutions of Ermakov-Pinney equation and equations of Painlev\'e II 
type.
The vector field $f(x)\frac{d}{dx} \in Vect(S^1)$ 
associated to the stabilizer orbit is called the projective 
vector field \cite{Hit} and the equation associated to this
is called the projective vector field equation $\Delta^{(3)}f = f_{xxx} + 4uf_x + 2u_xf = 0$. The operator 
$\Delta^{(3)} = \partial_{x}^{3} + 4u\partial_x + 2u_x$  associated to the projective vector field equation
equation is called the projective connection. In the next section we give a definition of projective connection
and its connection to the stabilizer set of the Virasoro orbit.

\smallskip 

In this paper we show an elegant geometrical connection between equations (\ref{1a}) and (\ref{1b})
in terms of higher-order projective connections $\Delta^{(n)}$ on $S^1$ \cite{Gu2,Gu3} and solution curves associated
to $\Delta^{(n)}f = 0$. 
Using the same kind of map we then construct two new negative KdV 
equations of Fuchssteiner type,
$$
\left(\frac{u_{txxx}}{3u^2 + u_{xx}}\right)_x + 10\left(\frac{(uu_t)_x}{3u^2 + u_{xx}}\right)_x + 3u_t = 0
$$
and
$$
\left(\frac{u_{txxxx}}{u^{\prime\prime\prime} + 16uu^{\prime}}\right)_x + 20\left(\frac{uu_{txx}}
{u^{\prime\prime\prime} + 16uu^{\prime}}\right)_x + 30\left(\frac{u_xu_{tx}}{u^{\prime\prime\prime} + 16uu^{\prime}}\right)_x
 + 18\left(\frac{u_{xx}u_{t} + 64u^2u_t}{u^{\prime\prime\prime} + 16uu^{\prime}}\right)_x + 4u_t = 0,
$$
and show these are associated to some equations of Qian type, $(- \frac{\psi_{xx}}{\psi})_t  = (\psi^3)_x$
and $(- \frac{\psi_{xx}}{\psi})_t  = (\psi^4)_x$, respectively.

\smallskip

 The projective
connections appear most naturally in 2-D conformal field theory (CFT) and integrable systems.
In 2-D CFT there exist several differential operators of various orders
which transform covariantly under the coadjoint action of $Diff(S^1)$.
Moreover, at least for $n \leq 4$, all these operators
depend only upon $u$ and its derivatives. These are also known as the Adler-Gelfand-Dikii (or AGD) operators.
Mathieu  has listed
several extended conformal operators in \cite{Math} and some of the members of this family are
$
\Delta^{(0)} = 1, \, \Delta^{(1)} = \partial_x,\, \Delta^{(2)} = \partial_{x}^{2} + u, \,
\Delta^{(3)} = \partial_{x}^{3} + 4u\partial_x + 2u^{\prime},
$
where $\Delta^{(2)}$ is the famous Hill's operator and $\Delta^{(3)}$ is the second Hamiltonian operator
for the KdV equation. Notice that the $\Delta^{(3)}$ operator also plays an imporatnt role in the inverse or negative 
KdV equation. In this article we focus on the next two higher-order operators, 
\beqn\label{1e}
\Delta^{(4)} \,=\,\partial_{x}^{4} + 9u^2 + 3u^{\prime\prime} + 
10u^{\prime}\partial_x + 10u\partial_{x}^{2}
\eeqn
and
\beqn\label{1f}
\Delta^{(5)} \,=\,\partial_{x}^{5} + 20u\partial_{x}^{3} + 30u^{\prime}\partial_{x}^{2} + 18u^{\prime \prime}\partial_x 
+ 64u^2\partial_x + 4u^{\prime\prime\prime} + 64uu^{\prime}. 
\eeqn
One must note that the operator $\Delta^{(5)}$ can be rescaled to
$$
{\tilde \Delta}^{(5)} \,=\,\partial_{x}^{5} + 10u\partial_{x}^{3} + 15u^{\prime}\partial_{x}^{2} + 9u^{\prime \prime}\partial_x 
+ 16u^2\partial_x + 2u^{\prime\prime\prime} + 16uu^{\prime}\eqno(6a)
$$
for $ u \to u/2$. We work with the second version of $\Delta^{(5)}$ operator.

\bigskip

The paper is organized as follows. In Section 2 we firstly introduce the stabilier set of the Virasoro orbit
and projective connection on $S^1$ and then we give the derivations of the equations of negative KdV
type. We establish geometrically the correspondences between the negative KdV equation of Qiao type and the KdV equation of
Fuchssteiner type. Using this geometrical correspondence we derive two 
new negative KdV equations in Section 3. In Section 4 we study the Painlev\'e properties and
symmetry analysis of these two newly found equations. We obtain soliton solutions of these
new equations in Section 5.

\section{Stabilizer Set of Virasoro orbit, projective connection on $S^1$ and negative KdV equation} 

Initially we give a description of the stabilier set of the Virasoro algebra and projective connection on $S^1$.
Then we explore their roles in the derivation of the equations of negative KdV type. In particular
we describe projective connections on $S^1$ and show their
roles for the construction of the KdV equations of Qian and Fuchssteiner type.

\smallskip

We consider the Lie algebra of vector fields on $S^1$, $Vect(S^1)$.
The dual of this algebra is identified with the
space of quadratic differential forms $u(x)dx^{\otimes2}$ 
by the pairing
$$  < u(x), f(x) >  = \int_{0}^{2\pi} u(x)\ f(x) dx,  $$
where $f(x) {d}/{dx}  \in  Vect(S^1).$
  The Virasoro algebra $Vir$ has a unique nontrivial 
  central extension by means of ${\bf R}$
 $$ 0 \longrightarrow {\bf R} \longrightarrow Vir
 \longrightarrow Vect(S^1) $$ 
described by the Gelfand-Fuchs cocycle 
$\omega_1(f,g) = \frac{1}{2} \int_{S^1} f^{\prime}g^{\prime \prime}dx.$

\bigskip

The elements of $Vir$ can be identified with the pairs 
($2\pi$ periodic function, real number).
The commutator in $Vir$ takes the form
$$ 
\left[ \left(f(x)\frac{d}{dx}, a\right), \left(g(x)\frac{d}{dx}, b\right) \right] 
= \left((fg^{\prime} - gf^{\prime})\frac{d}{dx}, 
\int_{S^1}\frac{1}{2} f^{\prime}g^{\prime \prime}\; dx\right). 
$$
The dual space ${Vir}^{\ast}$ can be identified with the
set $\{ (\mu, udx^2)~ |~ \mu \in {\bf R}$.

A pairing between a point $(\lambda, f(x){d}/{dx}) \in {Vir}$
and a point $\left(\mu, udx^{2}\right)$ is given by $\lambda \mu + \int_{S^1}f(x)u(x)\; dx.$

\bigskip

\begin{lem}
\beqn
 ad_{\left(\lambda, f(x)\frac{d}{dx}\right)}^{\ast} \left(\mu, udx^2\right) 
~=~ \frac{1}{2}\mu f^{\prime \prime \prime}
 + 2f^{\prime}u + 2fu^{\prime}. 
 \eeqn
 \end{lem}

{\bf Proof}: 
 It follows from the definition
 \begin{eqnarray*}
 < ad_{(\lambda, f)}^{\ast} (\mu, u), (\nu, g) > &&=~
 \left< (\mu, u), ad_{(\lambda,\; f)}(\nu, g) \right> \\
&& =~ \left< \left(\mu, u\right), \left(\frac{1}{2}\int_{S^1} 
f^{\prime}g^{\prime \prime} dx, \left[f\frac{d}{dx}, g\frac{d}{dx}\right]\right) \right> \\
&& =\int_{S^1}u\left(fg^{\prime} - f^{\prime}g\right)dx + 
\frac{1}{2}\mu \int_{S^1}f^{\prime}g^{\prime \prime}.
\end{eqnarray*}
$\Box$

\smallskip

\begin{cor}
The stabilizer space of the action of $f\frac{d}{dx} \in Vect(S^1)$ 
on the space of third-order differential operators of special type is given by
\beqn  f^{\prime \prime \prime} + 2u^{\prime}f + 4uf^{\prime} = 0, \eeqn
or
\beqn
ff^{\prime \prime} + 2uf^2 - \frac{1}{2}(f^{\prime})^2 = c,
\eeqn
where $c$ is a constant.
\end{cor}

\smallskip

\begin{dfnprop}
A vector field $v = f(x){d}/{dx}$ is a called projective 
vector field which keeps
fixed a given projective connection $\Delta = \frac{d^2}{dx^2} + u(x)$
\beqn  {\cal L}_v {\Delta}s ~=~ \Delta ({\cal L}_v s), \eeqn
for all $ s \in \Gamma (\Omega^{-\frac{1}{2}})$,
where ${\cal L}_v$ is the Lie derivative of $v$. A projective vector field $ v =  f{d}/{dx} \in \Gamma (\Omega^{-1})$
satisfies $$ f^{\prime \prime \prime} + 4f^{\prime}u + 2fu^{\prime} ~=~ 0.$$
\end{dfnprop}

\subsection{ Higher-order projective connections and Fuchssteiner's negative KdV equation}

We start with the definitions of projective connections.

\begin{dfn}[Projective Connection]
An extended projective connection on the circle is a class
of differential (conformal) operators, 
$$
\Delta^{(n)} : \Gamma (\Omega^{-\frac{n-1}{2}}) \longrightarrow 
\Gamma (\Omega^{\frac{n+1}{2}}), 
$$
such that \\
\noindent
 (1) \hbox{    } the symbol of $\Delta^{(n)}$  is 
 the identity, \\
\noindent
$$
(2)\hspace{30mm}  \hbox{      } \int_{S^1} (\Delta^{(n)} s_1)s_2 
~=~ \int_{S^1} s_1 (\Delta^{(n)} s_2) \hspace{45mm}
$$  
 for  all $ s_i \in \Gamma (\Omega^{-\frac{n-1}{2}}). $
\end{dfn}

\smallskip

It is known that the symbol of an $n$th-order 
operator from a vector bundle $U$ to $V$
is a section of $\hbox{ Hom  }(U, V \otimes {Sym^n T})$, where
$$ U = \Omega^{-\frac{(n-1)}{2}} \hbox{   } V = \Omega^{\frac{n+1}{2}}. $$
Because $T = \Omega^{-1}$, we have
$$ V \otimes Sym^{n}T \cong U, $$
thereby giving an invariant meaning to the first condition.

If $s_2 \in \Gamma (\Omega^{-\frac{n-1}{2}})$, then 
$s_1\Delta^{(n)}s_2 \in \Gamma (\Omega)$ is a one-form to integrate.

\smallskip

The consequence of the {\it first condition} is that all the differential
operators are monic, that is, the coefficient of the 
highest derivative is always
one. The {\it second condition} says that the term $u_{n-1} = 0$.  

\smallskip

The weights, $-\mbox{$\frac{1}{2}$}(n-1)$ and $\mbox{$\frac{1}{2}$}(n+1)$, related to the space
of operators $\Delta^{(n)}$ are known to physicists and mathematicians [19,37]
but not from the point view of projective connections.
 Consider a one-parameter family of $Vect(S^1)$ acting on the space
 of smooth functions $a(x) \in C^{\infty}(S^1)$ [31]
 \beqn\label{2a}
 {\cal L}_{v}^{\lambda}a(x) := f(x)a^{\prime}(x) - \lambda f^{\prime}(x)a(x),
 \eeqn
 where  ${\cal L}_{v}^{\lambda}$ is the Lie derivative with respect to
 $ v = f(x) \frac{d}{dx} \in Vect(S^1)$, given by
  \beqn\label{2b}
{\cal L}_{v}^{\lambda} := f(x)\frac{d}{dx} - \lambda f^{\prime}(x).
 \eeqn
 
\begin{dfn} 
The action of $Vect(S^1)$ on the space of Hill's operator
$\Delta \equiv \Delta^{(2)}$ is defined by the commutator with
the Lie derivative
\beqn\label{2c}
[{\cal L}_v , \Delta] := {\cal L}_{v}^{-3/2} \circ \Delta  -  
\Delta \circ {\cal L}^{1/2}.
\eeqn
\end{dfn}
 
This action can be identified with the 
coadjoint action of Virasoro algebra on its dual. 
Here we discuss this briefly.

Similarly we can generalize this action
on $\Delta^{(n)}$.  The $Vect(S^1)$ action on $\Delta^n$ is defined by
\beqn\label{2d}
[{\cal L}_v, \Delta^{(n)}] := {\cal L}_{v}^{-(n+1)/2} \circ 
\Delta^{(n)} - \Delta^{(n)} \circ 
{\cal L}_{v}^{(n-1)/2}.
\eeqn

\begin{prop}
A vector field is called a projective vector field if it keeps
fixed a given projective connection $\Delta$,
$$  {\cal L}_v {\Delta^{(n)}}s ~=~ \Delta^{(n)} ({\cal L}_v s), $$
for all $ s \in \Gamma (\Omega^{-\frac{n-1}{2}})$. It satisfies
\beqn\label{2e}
f^{\prime \prime \prime} + 4f^{\prime}u + 2fu^{\prime} ~=~ 0. 
\eeqn
\end{prop}

\bigskip

\noindent{\underline{\bf Illustration}}\\
\noindent
{\bf n = 3:} In this case the projective connection is defined by
$$ \Delta^{(3)} = \partial_{x}^{3} + 4u\partial_x + 2u^{\prime}
$$
and $s \in \Gamma(\Omega^{-1})$. It is easy to check that ${\cal L}_v \Delta^{(3)} = \Delta^{(3)} {\cal L}_v$
yields
$$
(f^{\prime \prime \prime} + 4f^{\prime}u + 2f u^{\prime} )^{\prime} \phi + 
(f^{\prime \prime \prime} + 4f^{\prime}u + 2f u^{\prime} )\phi^{\prime}
= 0. $$

\smallskip

\noindent{\bf n = 4:} The $4$th-order projective connection  
$\Delta^{(4)} = \partial_{x}^{4} + 9u^2 + 3u^{\prime \prime} + 
10u^{\prime}\partial_x + 10u\partial_{x}^{2}$ maps 
$\Delta^{(4)} : \Gamma(\Omega^{-\frac{3}{2}}) \longrightarrow 
\Gamma(\Omega^{\frac{5}{2}})$ and this immediately yields
$$
(\partial_{x}^{2} + 6u)(f^{\prime \prime \prime} + 4f^{\prime}u + 
2fu^{\prime}) + 5(f^{\prime \prime \prime} + 4f^{\prime}u + 2fu^{\prime} )^{\prime}
\phi^{\prime} + (f^{\prime \prime \prime} + 4f^{\prime}u + 
2fu^{\prime} )\phi^{\prime \prime} = 0.
$$
A similar result is found for other higher-order projective connections.

\bigskip

\begin{lem}
If $\psi_1$ and $\psi_2$ are solutions of Hill's equation
  \beqn\label{2f}
 \Delta \psi ~=~ \left(\frac{d^2}{dx^2} + u \right)\psi ~=~ 0, 
  \eeqn 
then the product $\psi_i \psi_j \in {\Gamma}(\Omega^{-1})$ satisfies
equation $ f^{\prime \prime \prime} + 2u^{\prime}f + 4uf^{\prime} = 0$
and traces out a three-dimesional space of solution.
\end{lem}

\noindent
{\bf Remark} The sections of $\Gamma (\Omega^{-{\frac{1}{2}}})$ which satisfy 
equation (\ref{2f}) are not functions, but the square root of a {\sl projective} 
vector field, because $ \psi \in \Omega^{- 1/2} $, the space of scalar densities 
of weight $ - 1/2 $,  is the square root of $f \in Vect(S^1 )$. 

\smallskip

\begin{prop}
Let $\psi$ be a solution of Hill's equation. If the flow on the immersion space satisfies 
$(- \frac{\psi_{xx}}{\psi})_t = (\psi^2)_x$, then
the stabilizer set of the Virasoro orbit
 satisfies the negative KdV equation of Fuchssteiner's type, namely 
$$\left(\frac{u_{txx}}{u_x}\right)_x + 4\left(\frac{uu_{t}}{u_x}\right)_x + 2u_t = 0.$$
\end{prop}

{\bf Proof} Because $\psi$ satisfies Hill's equation, we find that $u_t = (\psi^2)_x$. 
We know that, if $\psi$ is a solution of the
Hill equation, $\psi^2$ satisfies $ f^{\prime \prime \prime} + 2u^{\prime}f + 4uf^{\prime} = 0$.
Thus we obtain our desired result. $\Box$

\subsubsection{Connection with the nonholonomic deformed KdV equation}
In an interesting paper Kupershmidt \cite{Kup} constructed a nonholonomic 
deformation of the KdV equation.
By rescaling $v$ and $t$ he further modified this to
\begin{eqnarray}
u_t - 6uu_x - u_{xxx} + w_x \equal 0, \nonumber \\
 w_{xxx} + 4 uw_x + 2 u_x w \equal 0. 
\end{eqnarray}
This can be converted into bi-Hamiltonian form
\begin{equation}\label{2g}
u_t  = \ \ {\cal O}^2 
\bigg(\frac{\delta H_n}{\delta u}\bigg) 
- {\cal O}^1 (w), \ \ {\cal O}^2 (w) = 0, 
\end{equation}
where
\begin{equation}\label{2h}
{\cal O}^1 = \partial = \partial_x, \ \ \ {\cal O}^2 = \partial^3 + 2 (u \partial + \partial u) 
\end{equation}
are the two standard Hamiltonian operators of the $KdV$ hierarchy, $n = 2$, and
$H_1 = u, \ H_2 = u^2/2, \ H_3 = u^3/3 - u_x^2/2, \cdots $
are the conserved densities.
It is known that
the KdV6 equation always appears as a pair of equations, an evolution
equation of $u$ and a constraint equation of $w$.
In \cite{Gu4} we have studied various equivalent forms of the
nonholonomic deformation of the KdV equation. \bigskip

Equations (\ref{1a}) and (\ref{1b}) follow from the reduction of the 
nonholonomic deformation of the KdV equation.
If we assume $H_n = 0$ and set $w = -\psi^2$, 
then the evolution $u$ becomes $u_t = (\psi^2)_x$
and the constraint equation becomes
$(\psi^2)_{xxx} + 4u(\psi^2)_x + 2u_x(\psi^2) = 0, $
which in turn is related to the negative KdV equation for $u_t = (\psi^2)_x$.

\section{A new flow on immersion space and generalized flows of negative KdV type}

We define $n$ independent solutions
$(\psi_1, \psi_2, \cdots , \psi_n)$. The map 
\beqn\label{3a}  
x \longmapsto (\psi_1(x), \psi_2(x), \cdots , \psi_n(x) ), \qquad {\bf R} \longrightarrow {\bf R}P^{n-1},
\eeqn
defines an immersion in homogeneous cooordinates.

\begin{lem}
There is a one-to-one correspondence between\\ 
(1) the $n$th-order equation on $S^1$ 
$\Delta^{(n)}\psi = 0,$
where $\psi$ is the unknown function, and \\
(2) smooth orientation-preserving immersions $g : S^1 \longrightarrow {\bf R}P^{n-1}$,
modulo the equivalence upto $PSL(n,{\bf R})$.
\end{lem}

This {\it proof} goes as follows.  Given the $n$ independent solutions, $\psi_1, \cdots, \psi_2$, of 
the equation $\Delta^{(n)}\psi = 0,$
then $x \longmapsto (\psi_1 (x), \psi_2 (x), \cdots , \psi_n(x))$
defines a curve in the projective space
${\bf R}P^{n-1}$. Because the Wronskian
of the solution curve is constant upto multiplication by a matrix in $SL(n,{\bf R})$,
 then the Wronskian
of any immersion can be expressed by one. 
$\Box$

\bigskip

Thus we obtain a solution
curve associated to $\Delta^{(n)}$.  
As the coefficients are periodic, hence, if $\psi (x)$ is a solution, then
$\psi (x + 2\pi)$ is also a solution.  This implies that
$$ \psi (x + 2\pi) = M_{\psi} \psi (x), $$
where 
$$ M_{\psi} = \psi(2\pi){\psi(0)}^{-1} $$
is a monodromy matrix.  This matrix preserves the skew form given by the Wronskian so that 
$ det(M_{\psi}) = 1$, i.e., $M_{\psi} \in SL(n,{\bf R})$. If one chooses a different
solution curve, then a new monodromy matrix appears.  This is the conjugate
of $M_{\psi}$ by an element of $SL(n,{\bf R})$. This means that for each Lax operator
we can associate a projective curve the monodromy of which is an element of
the conjugacy class $[M_{\psi}]$. This curve is unique up to the projective action of
$SL(n, {\bf R})$.

\bigskip

\begin{lem}
Let $\psi_1$ and $\psi_2$ be solutions of Hill's equation. 
The equation 
\beqn\label{3b}
(a) \qquad f^{\prime\prime\prime\prime} + 10uf^{\prime\prime} 
+ 10u^{\prime}f^{\prime}
+ (9u^2 + 3u^{\prime\prime})f = 0
\eeqn
traces out a four-dimensional space of solutions spanned by
$\{\psi_{1}^{3},\, \psi_{1}^{2}\psi_2,\, \psi_{1}\psi_{2}^{2},\, \psi_{2}^{3}\}.
$ 
\beqn\label{3c}
(b) \qquad f^{\prime\prime\prime\prime\prime} + 20uf^{\prime\prime\prime} 
+ 30u^{\prime}f^{\prime\prime} + 18u^{\prime\prime}f^{\prime} + 64u^2f^{\prime} + 4u^{\prime\prime\prime}f +
+ 64uu^{\prime}f = 0
\eeqn
traces out a five-dimensional space of solutions spanned by
$\{\psi_{1}^{4},\, \psi_{1}^{3}\psi_2,\, \psi_{1}^{2}\psi_{2}^{2},\, \psi_{1}\psi_{2}^{3},\, \psi_{2}^{4} \}.
$ 
\end{lem}

{\bf Proof}: By direct lenghthy computation.
$\Box$

\bigskip

\begin{prop}
Let $\psi$ be a solution of Hill's equation $\psi_{xx} + u\psi = 0$.
\begin{enumerate}
\item Suppose that $\psi$ satisfies
the flow equation
\begin{equation}
\left(- \frac{\psi_{xx}}{\psi}\right)_t = 3\psi^2 \psi_x = \left(\psi^3\right)_x. \label{24a}
\end{equation}
Then $u$ satisfies
\begin{equation}
\left(\frac{u_{txxx}}{3u^2 + u_{xx}}\right)_x + 10\left(\frac{(uu_t)_x}{3u^2 + u_{xx}}\right)_x + 3u_t = 0.\label{24b}
\end{equation}
\item If $\psi$ satisfies
\begin{equation}
\left(- \frac{\psi_{xx}}{\psi}\right)_t = 4\psi^3 \psi_x =  \left(\psi^4\right)_x, \label{25a}
\end{equation}
then $u$ satisfies
\begin{equation}
\left(\frac{u_{txxxx}}{u^{\prime\prime\prime} + 16uu^{\prime}}\right)_x + 20\left(\frac{uu_{txx}}
{u^{\prime\prime\prime} + 16uu^{\prime}}\right)_x + 30\left(\frac{u_xu_{tx}}{u^{\prime\prime\prime} + 16uu^{\prime}}\right)_x
 + 18\left(\frac{u_{xx}u_{t} + 64u^2u_t}{u^{\prime\prime\prime} + 16uu^{\prime}}\right)_x + 4u_t = 0\label{25b}.
\end{equation}
\end{enumerate}
\end{prop}

{\bf Proof}  Because $\psi$ satisfies Hill's equation, hence (\ref{24a}) yields $u_t = \left(\psi^3\right)_x$. 
Substituting this expression into (\ref{3b}) we obtain our result. Similarly, when we substitute $u_t = (\psi^4)_x$ into (\ref{3c}), 
we obtain (\ref{25b}).
$\Box$

\section{Painlev\'e and symmetry analyses for the newly derived equations}

The equation
\begin {equation}
\left (\frac {u_{txx}} {u_x}\right)_x + 4\left (\frac {uu_t} {u_x}\right)_x + 2u_t = 0, \label {6.1}
\end {equation}
in which $u = u (t,\,x) $, possesses the symmetries\footnote{Courtesy of the mathematica add-on Sym 
\cite{Dimas 05 a, Dimas 06 a, Dimas 08 a}.}
\begin {eqnarray}
& & \Gamma_1 = \partial_x, \label {6.2} \\
& & \Gamma_2 = x\partial_x - 2u\partial_u \quad \mbox {\rm and} \label {6.3} \\
& & \Gamma_3 = a (t)\partial_t.  \label {6.4}
\end {eqnarray}
The function $a(t) $ is arbitrary, apart from the requirement that it be differentiable, and reflects the fact that 
(\ref {6.1}) is homogeneous in the first derivative with respect to time.

\strut\hfill

The algebra is $A_2\oplus A_{\infty} $.

Equation (\ref{24b}) can be expanded by the denominator removed to give
\begin {eqnarray}
& & 27 u^4 u_{ t}-60 u u_{ x}{}^2 u_{ t}+48 u^2 u_{ xx} u_{ t}+13 u_{ xx}{}^2 u_{ t}-10 u_{ x} u_{ xxx} u_{ t} \nonumber \\
&& +20u_{ x} u_{ xx} u_{ tx} -10 u u_{ xxx} u_{ tx}+30 u^3 u_{ txx}+10 u u_{ xx} u_{ txx} \nonumber \\
&& -6 u u_{ x} u_{ txxx}-u_{ xxx} u_{txxx}+3 u^2 u_{txxxx}+u_{xx} u_{txxxx} = 0. \label {101}
\end {eqnarray}
To seek a travelling-wave solution of (\ref {101}) we make the substitution $u (t,\, x) \longrightarrow w (v) $, where $v = x - ct $.  The fifth-order equation subsequent upon this substitution  and division by $-c $ is
\begin {eqnarray}
&& 27 w^4 w'-60 w w'^3+48 w^2 w' w''+33 w' w'' {} ^2+30 w^3 w'''-10 w'^2 w''' - 6 w w' w'''' \nonumber \\
&& -w''' w'''' + 3 w ^2 w^{(5)}+w'' w^{(5)} = 0. \label {102}
\end{eqnarray}
(Note that this equation replaces (23) in the text.)

\strut\hfill

Equation (\ref {102}) has just the two symmetries $\Gamma_1 = \partial_v $ and $\Gamma_2 = 2v\partial_v - 3w\partial_w $ with the algebra $A_2 $ in the Mubarakzyanov Classification Scheme \cite{Morozov 58 a, Mubarakzyanov 63 a, Mubarakzyanov 63 b, Mubarakzyanov 63 c}.  The number of Lie point symmetries is insufficient to reduce the equation completely.

\strut\hfill

We investigate (\ref {102}) from the point of view of singularity analysis.  The exponent of the leading-order term is $-2 $ and the equation for the coefficient is the solution of
\begin {equation}
-1440  a^2 - 2136 a^3 - 816 a^4 - 54 a^5 = 0, \label {103}
\end {equation}
namely $a = -12,\, -2\,\mbox {\rm and}\, -10/9 $ in addition to a double zero.  The equation to be satisfied by the resonances is
\begin {eqnarray}
&& -2880 a-6408 a^2-3264 a^3-270 a^4+3048 a s+3712 a^2 s+828 a^3 s \nonumber \\
&& +27 a^4 s-1416 a s^2 -1320 a^2 s^2-366 a^3 s^2+474 a s^3+257 a^2 s^3+30a^3 s^3 \nonumber \\
&& -96 a s^4-48 a^2 s^4 +6 a s^5+3 a^2 s^5 = 0. \label {104} 
\end{eqnarray}
The resonances corresponding to the three values of the coefficient of the leading-order term are listed in Table 1.
\begin {table}
\begin {center}
\caption {Listing of the resonances corresponding to the three possible values of the coefficient of the leading-order term (\ref{24a})}
\[
\begin {array} {|r | |c |}
\hline
\mbox {\rm Coefficient} & \mbox {\rm Set of Resonances} \\
 & \\
\hline
-12 & -1,\, 2,\, 14,\,\half\left (1 - \sqrt {337}\right),\,\left (1 + \sqrt {337}\right) \\
 & \\
-2 & -2,\, -1,\, 6 \\
 & \\
-\frac {10} {9} &  -1,\, 2,\,\frac {10} {3},\,\frac {14} {3},\, 7 \\
 & \\
\hline
\end{array}
\]
\end {center}
\end {table}

\strut\hfill

Some observations are in order.  Firstly only the third resonance, $-10/9 $, has the full number of acceptable resonances.  The first resonance, $-12 $ certainly has five resonances, but two of them are quite irrational.  The second coefficient does not even have the right number of resonances.  The Check Sum for the first and third resonances is 22 whereas that for the second resonance is 3.  The Laurent expansion is a Right Painlev\'e Series in $(v - v_0) ^ {1/9} $.  Note that the expansion would be valid only on a section of the punctured disc due to the presence of powers which are multiples of 1/9.

\strut\hfill

There has been no success in attempts to integrate the equation directly.

\strut\hfill

Equation (\ref{25b}) has the travelling-wave form
\begin {eqnarray} 
 & & 9456 w^2 w'^3-768 w'^3 w''+3200 w w'^2 w'''+832 w^2 w'' w'''+48 w''{}^2 w''' \nonumber \\
 &&+72 w' w'''{}^2-832 w^2 w' w^{ (4) }-48 w'w'' w^{ (4) }-16 w'^2 w^{(5)} \nonumber \\
& & -16 w w'' w^{(5)}-w^{ (4) } w^{(5)}+16w w' w^{(6)}+w'' w^{(6)} = 0.  \label {105}
\end{eqnarray}
Again the exponent of the leading-order term is $-2 $.  The coefficient of the leading-order term satisfies the equation
\begin {equation}
-34560 a^2-101376 a^3-190464 a^4-155648 a^5 = 0.  \label {106}
\end {equation}
The roots of (\ref {106}) are given by
\begin {equation}
\left\{a\to -\frac{3}{4}\right\},\,\{a\to 0\},\,\{a\to 0\},\,\left\{a\to \frac{3}{76} \left(-6-i \sqrt{154}\right)\right\},\,\left\{a\to \frac{3}{76} \left(-6+i \sqrt{154}\right)\right\}.  \label {107}
\end {equation}
The equation for the resonances is
\begin {eqnarray}
& & -69120 a-304128 a^2-761856 a^3-778240 a^4+87552 a s+221184 a^2 s \nonumber \\
&& +427520 a^3 s+233472 a^4 s-47136 a s^2-98176 a^2 s^2 \nonumber \\
& & -55808 a^3 s^2+16320a s^3+29056 a^2 s^3-5504 a^3 s^3-3960 a s^4-5664 a^2 s^4 \nonumber \\
&& +1664 a^3 s^4+528 a s^5+704 a^2 s^5-24 a s^6-32 a^2 s^6 = 0.  \label {108}
\end {eqnarray}
The resonances for the nonzero values of $a $ are in Table 2.
\begin {table}
\begin {center}
\caption  {Listing of the resonances corresponding to the three possible values of the coefficient of the leading-order term (\ref{24b})}
\[
\begin {array} {|r | |c |}
\hline
\mbox {\rm Coefficient} & \mbox {\rm Set of Resonances} \\
&  \\
\hline
-\frac {3} {4} & -2,\, -1,\, 4,\, 6 \\
& \\
 \frac{3}{76} \left(-6-i \sqrt{154}\right) & -1,\, 2,\, 6,\, 8, \\
&  \frac{1}{38} \left(133-\sqrt{19 \left(-917+72 i \sqrt{154}\right)}\right),\, \frac{1}{38} \left(133+\sqrt{19
\left(-917+72 i \sqrt{154}\right)}\right) \\
& \\
\frac{3}{76} \left(-6+i \sqrt{154}\right) & -1,\, 2,\, 6,\, 8,\, \\
&  \frac{1}{38} \left(133-\sqrt{19 \left(-917-72 i \sqrt{154}\right)}\right),\, \frac{1}{38} \left(133+\sqrt{19\left(-917-72 i \sqrt{154}\right)}\right) \\
& \\
\hline
\end{array}
\]
\end {center}
\end {table}
In the case of the first resonance the check sum is 7 whereas for the second and third resonances it is 22.

\strut\hfill

None of the possible values of the coefficient of the leading-order term gives a satisfactory set of resonances.  In the cases of the second and third one could opine that the complex values completely destroy any possible results from the singularity analysis.  The values of the resonances for the first coefficient are acceptable, albeit of insufficient number.  In the past there has been talk of possessing the Partial Painlev\'e Property, but the idea has not been accepted by the experts in the area of singularity analysis.

\section{Solitonic solutions of generalised negative KdV equations}

We consider the possibility of the existence of solitonic solutions to the three partial differential equations of interest.  The first is (\ref{24a}))
\begin {equation}
\left (\psi ^ 3 (t,\,x)\right)_x + \left (\frac {\psi (t,\,x)_{xx}} {\psi (t,\,x)}\right)_t = 0.  \label {7.1}
\end {equation}
We make the substitution
\begin {equation}
\psi (t,\,x) = A\, \sech ^n (x -pt), \label {7.2}
\end {equation}
where the three parameters $A $, $n $ and $p $ are to be determined.  After some simplification we have
\begin {equation}
2 A ^ 2 n (1+ n) p\,\sech ^ {3+2n} (x -pt) + 3 A ^ 5p\, \sech ^ {1+5n} (x -pt) = 0.  \label {7.3}
\end {equation}
We achieve balance by equating the two exponents which means that $n = 2/3 $. It then follows that the `speed of propagation' is given by $p = -9 A ^ 3/10 $.

\strut\hfill

The second equation (\ref{25a}) is
\begin {equation}
\left (\psi ^ 4 (t,\,x)\right)_x + \left (\frac {\psi (t,\,x)_{xx}} {\psi (t,\,x)}\right)_t = 0.  \label {7.4}
\end {equation}
We make the same substitution as in (\ref {7.2}) and the simplified equation is
\begin {equation}
p (1+n)\left (1-\,\tanh ^ 2 (x -pt)\right) + 2 A ^ 4\, \sech ^ {4n} (x -pt) = 0.  \label {7.5}
\end {equation}
Evidently balance is achieved by setting $n = \half $ and $p = -4 A ^ 4/3 $.  

\strut\hfill

The third equation is
\begin {equation}
\left (\frac {u (t,\,x)_{txx}} {u (t,\,x)}\right)_x + 4\left (\frac {u (t,\,x)u (t,\,x)_x} {u (t,\,x)_x}\right)_x + 2u (t,\,x)_t = 0.  \label {7.6}
\end {equation}
After the substitution of (\ref {7.2}) and some simplification we have
\begin {eqnarray}
&& 2 A ^2 n ^ 2p\, \sech ^ {3+2n} (x -pt)\, \sinh ^ 3 (x -pt)\left\{2+3n +n ^ 2+3 An\, \sech ^n (x-pt) \right. \nonumber \\
&& \qquad \left. +\left (2+3n +n ^ 2\right)\, \tanh ^ 2 (x-pt)\right\} = 0 \label {7.7}
\end {eqnarray}
in which balance is achieved by setting $n = 2 $ and, as a consequence, $A = 2 $.

\strut\hfill

So the three equations all have solitonic solutions.

\section{Conclusion}

We have given a brief description of the projective connections on $S^1$ and 
their role for the construction of generalised equations of negative KdV type.  Using our method we established a connection between the 
negative KdV equation of Fuchssteiner and that of Qiao.
In this paper we studied two equations of generalised Qiao type, namely,
$(- \frac{\psi_{xx}}{\psi})_t = 3\psi^2 \psi_x$ and $(- \frac{\psi_{xx}}{\psi})_t = 4\psi^3 \psi_x$
and mapped these two generalised negative KdV equations to
generalized equations of Fuchssteiner type. 
We have studied symmetries and Painlev\'e properties of these two latter equations. We also
showed that they admit solitonic solutions.

\section*{Acknowledgements}
It is our pleasant duty to acknowledge gratefully
for several stimulating discussions with Professor Valentin Ovsienko. 
We are grateful to Professors Z. Qiao and Andy Hone for their remarks and encouragement.
PGLL is indebted to SNBNCBS for excellent hospitality during his visits.

\end{document}